\documentclass[aps,prl,reprint,groupedaddress]{revtex4-1}
\usepackage{braket}
\usepackage[]{graphicx}
\usepackage{subfigure}
\usepackage[table]{xcolor}
\usepackage{amsbsy}
\usepackage{amsfonts}
\usepackage{MnSymbol}

\begin{document}

\title{Experimental measurement of the self-healing of the spatially inhomogeneous states of polarization of radially and azimuthally polarized vector Bessel beams }

\author{Giovanni Milione,$^{1,2,3*}$ Angela Dudley,$^{4*}$ Thien An Nguyen,$^{1,3}$ Ougni Chakraborty,$^{1,3}$ \\ 
Ebrahim Karimi,$^{5}$ Andrew Forbes,$^{4,6}$ Robert R. Alfano,$^{1,2,3}$}
\email[Electronic address: ]{giomilione@gmail.com, angelajvr@gmail.com}

\affiliation{$^1$Physics Department, City College of New York of the City University of New York, 160 Convent Ave. New York, NY 10031 USA}
\affiliation{$^2$Graduate Center of the City University of New York, 365 Fifth Ave., New York, NY 10016 USA}
\affiliation{$^3$New York State Center for Complex Light, 160 Convent Ave. New York, NY 10031 USA}
\affiliation{$^4$CSIR National Laser Centre, P.O. Box 395, Pretoria 0001, South Africa}
\affiliation{$^5$Department of Physics, University of Ottawa, 150 Louis Pasteur, Ottawa, Ontario, K1N 6N5 Canada}
\affiliation{$^6$School of Physics, University of the Witwatersrand, Johannesburg 2000, South Africa}
\affiliation{$^*$Corresponding authors: giomilione@gmail.com, angelajvr@gmail.com}

\date{\today}

\begin{abstract} We experimentally measured the self-healing of the spatially inhomogeneous states of polarization of radial and azimuthal polarized vector Bessel beams. Radial and azimuthal polarized vector Bessel beams were generated via a digital version of Durnin's method, using a spatial light modulator in concert with a liquid crystal $q$-plate. As a proof of principle, their intensities and spatially inhomogeneous states of polarization were measured using Stokes polarimetry as they propagated through two disparate obstructions. It was found, similar to their intensities, the spatially inhomogeneous states of polarization of a radial and azimuthal polarized vector Bessel beams self-heal. Similar to scalar Bessel beams, the self-healing of vector Bessel beams can be understood via geometric optics, i.e., the interference of the unobstructed conical rays in the shadow region of the obstruction. The self-healing of vector Bessel beams may have applications in, for example, optical trapping. 
\end{abstract}

\pacs{XXX, XXX}


\maketitle

\noindent A Bessel beam is a light beam that is a solution to the Helmholtz wave equation, existing over a limited region of propagation, and experimentally generated by the interference of conical rays \cite{Durnin1, Durnin2}. It possesses a property referred to as ``self-healing," i.e., its intensity reappears after propagation through an obstruction. Self-healing can be simply understood via geometric optics \cite{Litvin}. When a portion of a Bessel beam is obstructed in one plane, the unobstructed conical rays interfere in its shadow region in another plane, as shown in Fig. 1(e). Due to this property, Bessel beams have been extensively studied and have been used for a number of applications; for comprehensive reviews see \cite{Review1,Review2, Review3, Dudley1}. For example, when using a Bessel beam for optical trapping, it is possible to simultaneously trap multiple particles in well separated planes \cite{Dholakia}, and make particle tractor beams \cite{Dogariu, Novitsky1,Novitsky2,Ruffner}.

A vector beam is a light beam possessing a spatially inhomogeneous state of polarization such as radial or azimuthal polarization as shown in Fig. 2(b) and Fig. 2(c), respectively. Vector beams have received significant interest \cite{Zhan, Brown}, due in great part to their ability to produce stronger longitudinal field components \cite{Brown1, Brown2}, and smaller spot sizes \cite{Leuchs}, as compared to scalar light beams, upon focusing by high numerical aperture objectives. Also, when using vector beams for optical trapping, it is possible to improve the axial and transverse stiffness of the optical trap via radial and azimuthal polarization, respectively \cite{Dunlop, Toussaint, Sato}. Most recently, vector beams have been used for optical communication \cite{optica}.

Like any other light beam a Bessel beam can have a \textit{scalar} (spatially homogeneous) or \textit{vector} (spatially inhomogeneous) state of polarization \cite{Bouchal,Aiello}. The great majority of experimental investigations of Bessel beams concern scalar Bessel beams.  Yet, a vector Bessel beam possesses the properties of a Bessel beam and a vector beam, as described above, and may be used in comparable applications. For example, it may be possible to improve the axial and transverse stiffness of a tractor beam when using a vector Bessel beam. While there are extensive studies on self-healing of scalar Bessel beams, there are limited studies on self-healing of vector Bessel beams \cite{Hasman,Vyas, He, Wu}, particularly with respect to their spatially inhomogeneous states of polarization. Previous work only measured the propagation and self-healing of the \textit{intensities} of vector Bessel beams. 

	In this work, we experimentally measured the self-healing of the spatially inhomogeneous states of \textit{polarization} of radial and azimuthal polarized vector Bessel beams. Radial and azimuthal polarized vector Bessel beams were generated via a digital version of Durnin's method, using a spatial light modulator in concert with a liquid crystal $q$-plate. As a proof of principle, their intensities and spatially inhomogeneous states of polarization were measured using Stokes polarimetry as they propagated through two disparate obstructions. It was found, similar to their intensities, the spatially inhomogeneous states of polarization of a radial and azimuthal polarized vector Bessel beams self-heal. Similar, to scalar Bessel beams, the self-healing of vector Bessel beams can be understood via geometric optics, i.e., the interference of unobstructed conical rays in the shadow region of the obstruction. The self-healing of vector Bessel beams may have applications in, for example, optical trapping. 
\begin{figure*}[htb]
\centerline{\includegraphics[scale = .9]{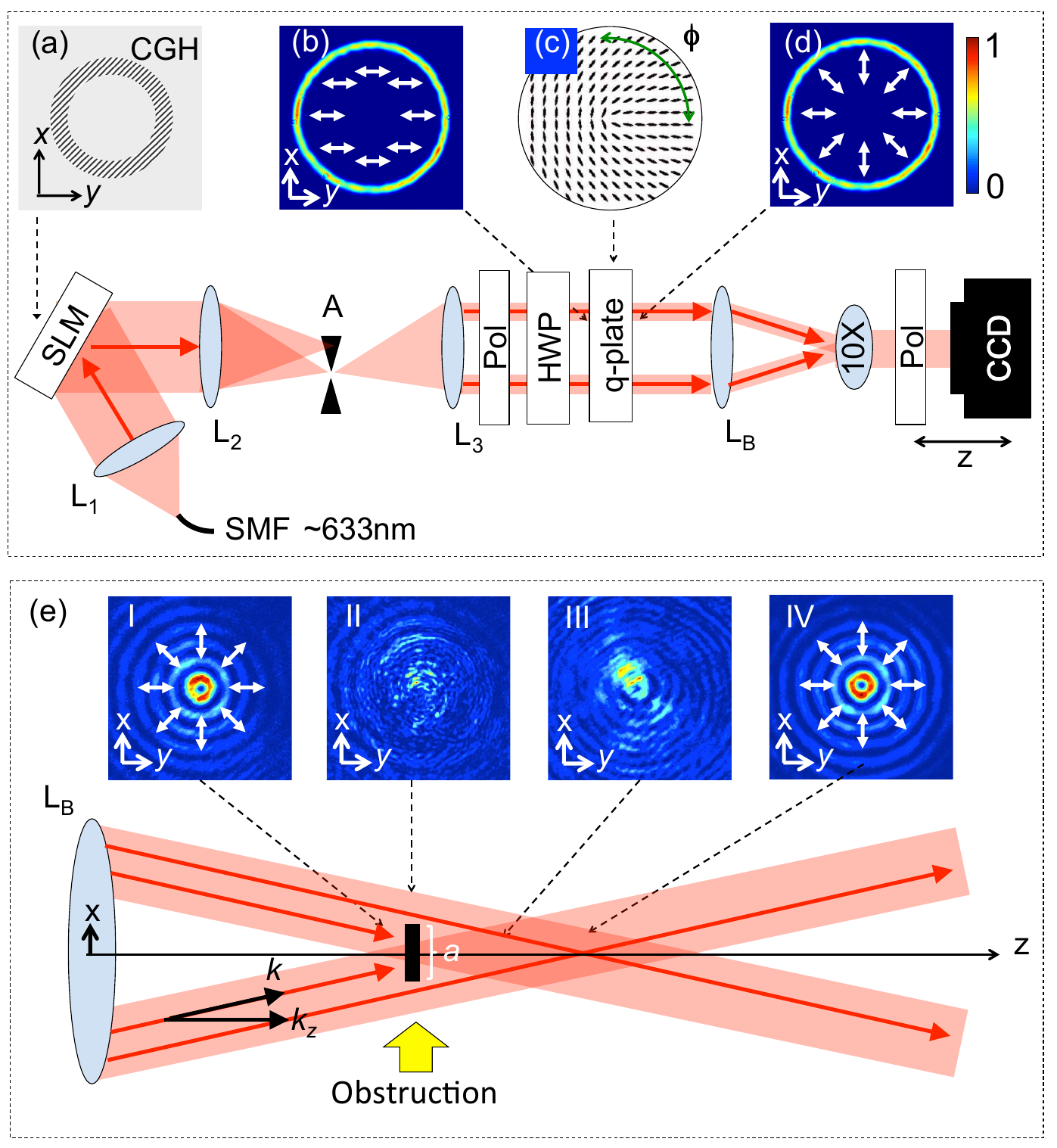}}
\caption{Experimental setup: (a) Computer generated hologram (CGH) of an annular slit additionally superimposed with a linear grating displayed on spatial light modulator( SLM). (b) Linear polarized annular ring of light before $q$-plate. (c) Schematic of $q=1/2$ $q$-plate (d) Radially polarized annular ring of light after $q$-plate. (e) Schematic of vector Bessel beam propagation though an obstruction in the focal region of $L_B$. The intensity of the vector Bessel beam is shown at a propagation distance of roughly (I) 64 mm (unobstructed), (II) 82 mm (obstructed), (III) 118 mm (semi-healed) and (IV) 136 mm (self-healed) after the Fourier transforming lens LB.}
\end{figure*}

A schematic of the experimental setup is shown in Fig. 1. First, a scalar Bessel beam was generated following Durnin's method. In Durnin's method, an annular slit is placed in the back focal plane of a lens and illuminated with a collimated light beam resulting in an annular ring of light.  Each point along the annular ring acts as a point source which the lens transforms into a good approximation to a Bessel beam in its focal region \cite{Durnin1, Durnin2}. A digital method of Durnin's method was implemented \cite{Forbes}. An annular slit, additionally superimposed with a linear grating, was created using a computer generated hologram (CGH) displayed on a reflective phase only spatial light modulator (SLM) (HoloEye) as shown in Fig. 1(a). A linear polarized HeNe laser beam ($\lambda \sim$ 633 nm) was spatially filtered by a single mode optical fiber (SMF), expanded, collimated by a lens ($L_1$), and then illuminated the SLM. The light at the plane of the SLM was then spatially filtered by an aperture (A) in the first diffraction order of a 4f imaging system ($L_2$ and $L_3$), imaged onto the back focal plane of a 10 cm focal length lens ($L_B$), resulting in the linear polarized annular ring of light shown in Fig. 1(b). A good approximation to a scalar (linearly polarized) Bessel beam was formed in the focal region of lens $L_B$. 
\begin{figure*}[htb]
\centerline{\includegraphics[scale = .9]{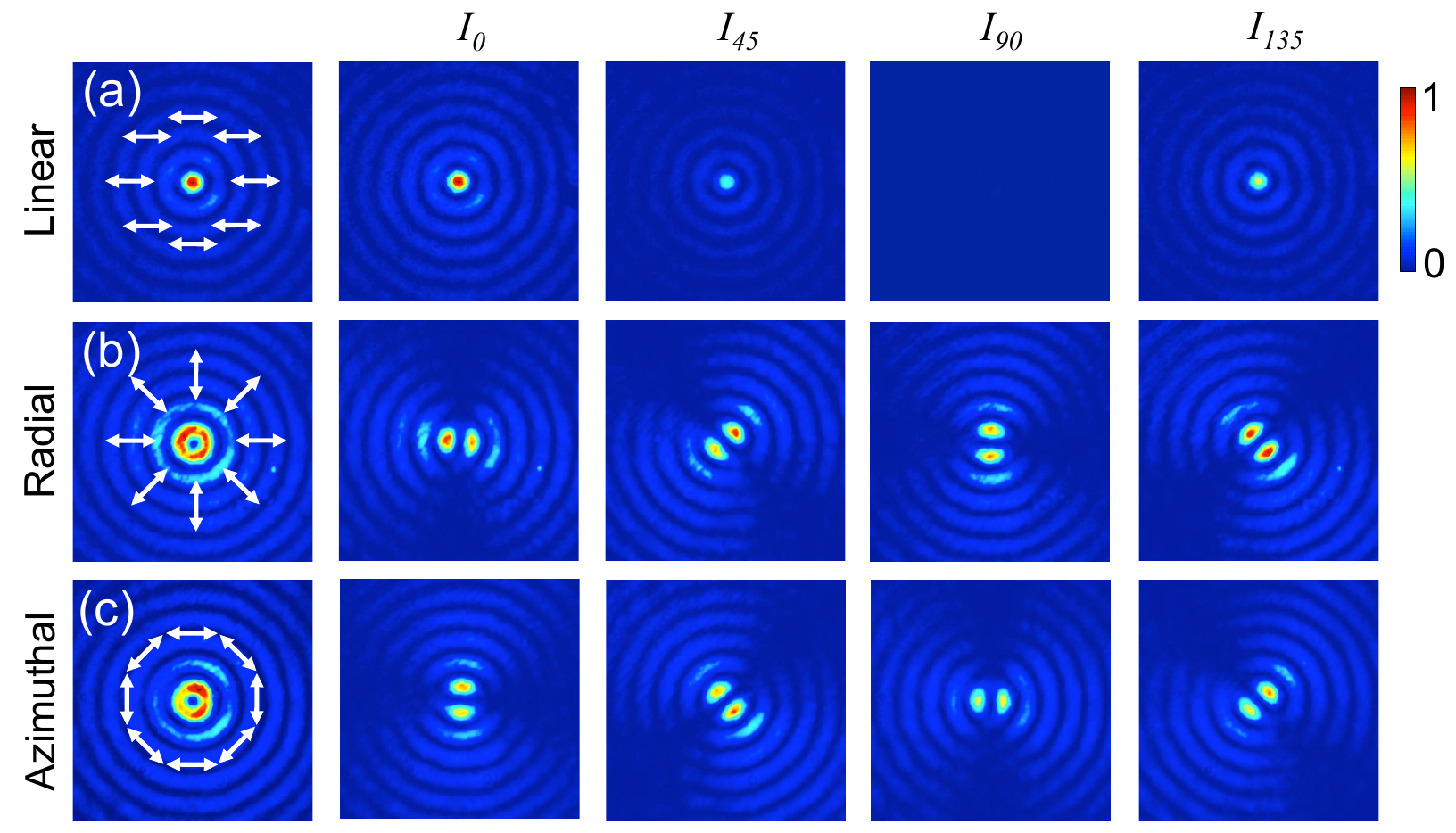}}
\caption{Experimentally measured intensity of scalar (linearly polarized) and vector Bessel beams. (a) (first row) Scalar Bessel beam. (b) (second row) Radial polarization. (c) (third row) Azimuthal polarization. The columns show $I_0$ (second column), $I_{45}$ (third column), $I_{90}$ (fourth column), $I_{135}$ (fifth column) for each Bessel beam as described in the text.}
\end{figure*}

Next, the scalar (linearly polarized) Bessel beam was converted into a vector Bessel beam. There are many methods to generate vector beams including the use of metasurfaces and optical fibers \cite{Lin,Milionea,Milioneb,Milionec}. Here, we use a $q$-plate \cite{Dudley}.  A $q$-plate is a liquid crystal technology comprising of a thin layer of liquid crystal molecules in-between two thin glass plates. The orientation of the liquid crystal molecules is described by $q \phi$, where $\phi$ is the azimuthal coordinate, and $q$ is a half integer. A $q=1/2$ $q$-plate is schematically shown in Fig. 1(c). Effectively, a $q$-plate is a half wave plate with an azimuthally varying fast axis that can be represented by the Jones matrix \cite{marrucci}:
\begin{equation}
\hat{Q} = \left(\begin{array}{cc} \cos 2q \phi & \sin 2 q \phi \\ \sin 2 q \phi & -\cos 2q  \phi \end{array}\right).
\end{equation}
Using Jones calculus, it can be easily shown, for a $q=1/2$ $q$-plate, horizontal (vertical) polarization can be converted into radial (azimuthal) polarization \cite{Karimi2}.  In general, a $q$-plate converts any state of polarization on the Poincare sphere to a higher-order state of polarization on the higher-order Poincare sphere \cite{Milione1,Milione2}. The \textit{q}-plate is also ``tunable"; the amount of the incident light's power converted to radial or azimuthal polarization is directly controlled, i.e., tuned, via a voltage over the $q$-plate. For $\lambda \sim 633$nm, when the voltage over the \textit{q}-plate is $V_o \sim $5 volts, no light will be converted to radial (azimuthal) polarization, i.e., the light remains linear polarized. When the voltage over the \textit{q}-plate is $V_v \sim$2.3  volts, all of the incident light's power will be converted to radial (azimuthal) polarization \cite{Slussarenko, Milione4}. 

The $q$-plate was placed close to the back focal plane of lens $L_B$.  A linear polarizer (Pol) was placed just before the $q$-plate to ensure the incident light was completely linear polarized. A half wave plate (HWP) was used to rotate the light's polarization horizontal (vertical).  A signal generator generating a 1kHz square wave was used to apply a voltage over the $q$-plate.  When the voltage was $V_o$, a good approximation to a scalar (linearly polarized) Bessel beam was generated in the focal region of lens $L_B$, as shown in Fig. 2(a). When the voltage was $V_v$, and the light's polarization was rotated horizontal (vertical), a good approximation to a radial (azimuthal) polarized vector Bessel beam was generated in the focal region of lens $L_B$, as shown in Fig. 2(b) (Fig. 2(c)). A 10X microscope objective was used to image each Bessel beam in the focal region of lens $L_B$ onto a CCD camera. 

Stokes polarimetry was used to measure the state of polarization of each Bessel beam \cite{Goldstein, Angela}. In Stokes polarimetry, the Stokes parameters are measured via intensity measurements and used to calculate the polarization orientation and polarization ellipticity at every spatial point of a light beam. The first three Stokes parameters are given by \cite{Goldstein, Angela}: 
\begin{eqnarray} 
S_0(r,\phi) &=& I_0(r,\phi) + I_{90}(r,\phi), \\ 
S_1(r,\phi) &=& I_0(r,\phi) - I_{90}(r,\phi), \\ 
S_2(r,\phi) &=& I_{45}(r,\phi) - I_{135}(r,\phi),
\end{eqnarray}
where $I_0(r,\phi)$, $I_{45}(r,\phi)$, $I_{90}(r,\phi)$, and $I_{135}(r,\phi)$ are the intensity of the light beam, at every spatial point, measured after a linear polarizer whose transmission axis is rotated $0^{\circ}$, $45^{\circ}$, $90^{\circ}$, $135^{\circ}$, respectively. $(r,\phi)$ are cylindrical coordinates. The total intensity of the light beam is given by  $S_0(r,\phi)$ and the orientation of the state of polarization is given by \cite{Goldstein, Angela}:
\begin{equation}
\psi(r,\phi) = \frac{1}{2} \tan^{-1}  \Big( \frac{S_2(r,\phi)}{ S_1(r,\phi) }  \Big).
\end{equation}
A linear polarizer (Pol) was placed just before the CCD camera in the experimental setup described above. The linear polarizer was used to measure $I_0(r,\phi)$, $I_{45}(r,\phi)$, $I_{90}(r,\phi)$, and $I_{135}(r,\phi)$ for each Bessel beam as shown in the respective columns of Fig. 2. While the third Stokes parameter, $S_3$, and therefore the polarization ellipticity, was not measured, as shown in the experimental results, it is qualitatively enough to visualize the self-healing of the spatially inhomogeneous state of polarization of each vector Bessel beam via the polarization orientation.

Finally, each Bessel beam was made to propagate through an obstruction. The results are shown in Fig. 3. The obstruction was created by placing a pitted glass slide possessing multiple, random, speckled, and opaque obstructions in the path of the Bessel beam. The slide's position was adjusted until an isolated and appropriately sized obstruction was found. The size and position of the obstruction was chosen such that it obstructed approximately a small portion of the Bessel beam near its center. The obstruction is outlined by a dashed white line in Fig. 3. The intensities and states of polarization of each Bessel beam were measured via Stokes polarimetry, as described above, at four propagation distances as they propagated through the obstruction. The four propagation distances are schematically shown in Fig. 1(e): (I) unobstructed (II) obstructed (III) semi-healed (IV) self-healed. $S_0(r,\phi)$ is overlaid with $\psi(r,\phi)$ such that any change in intensity and orientation  of the state of polarization as the Bessel beam propagates through the obstruction can be visualized simultaneously; $S_0(r,\phi)$ is encoded by relative contrast, i.e., bright to dark, and $\psi(r,\phi)$ is encoded by color. 
\begin{figure*}[htb]
\centerline{\includegraphics[scale = .9]{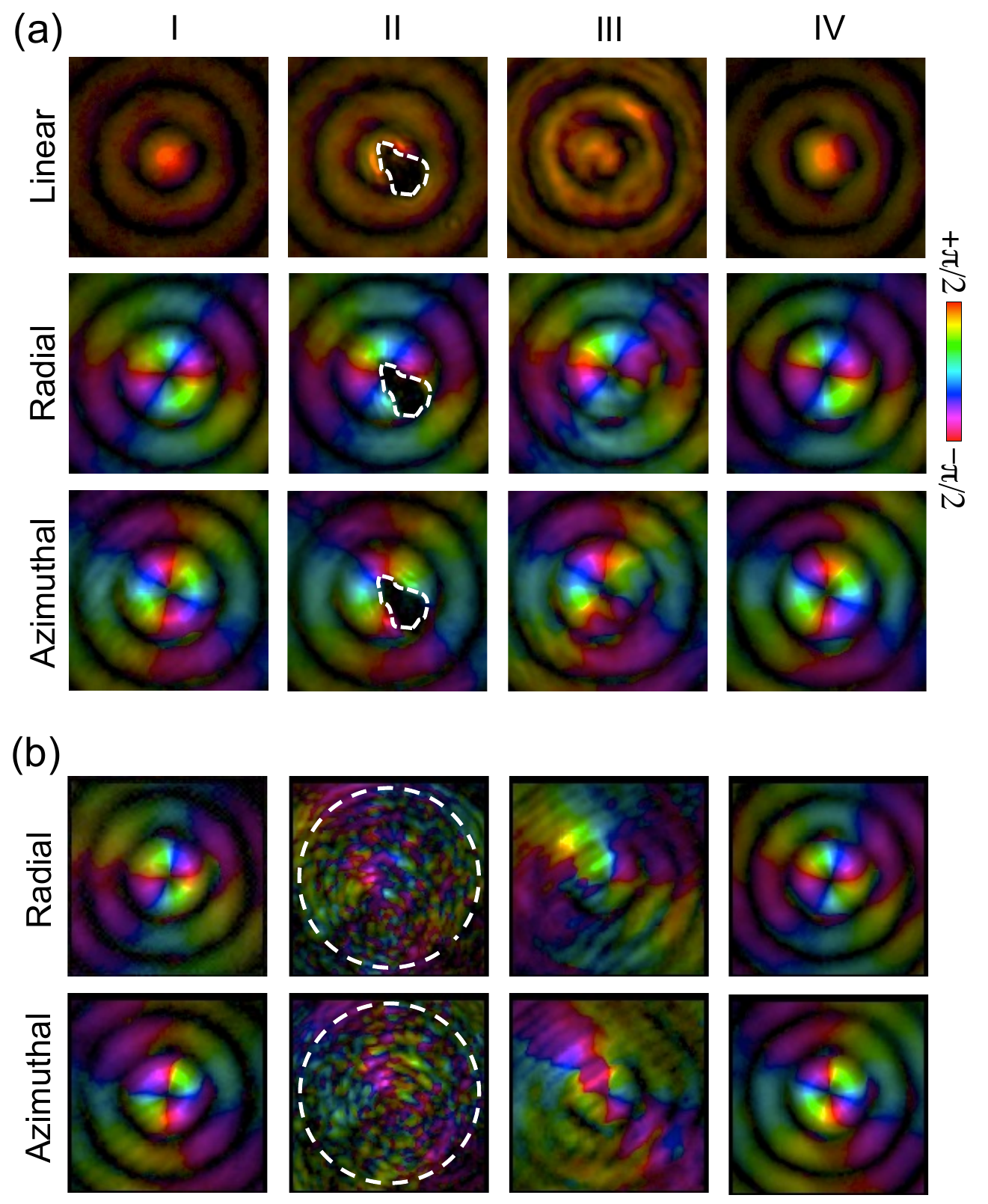}}
\caption{Experimentally measured intensities and spatially inhomogeneous states of polarization of scalar (linearly polarized) and vector Bessel beams as they propagated through two different obstructions at four different propagation distances. The obstruction is outlined by a white dashed line. (column I) Unobstructed. (column II) Obstructed. (column III) Semi-healed. (column IV) Self-healed. The Bessel beams' total intensities, $S_0(r,\phi)$, are encoded by relative contrast, and overlaid with the orientation of their states of polarization, $\psi(r,\phi)$, encoded by color. (a) First obstruction (first row) Scalar Bessel beam. (second row) Radial polarization (third row) Azimuthal polarization. (b) Second obstruction (first row) Radial polarization. (second row) Azimuthal polarization.}
\end{figure*}

The self-healing of the intensity and spatially inhomogeneous state of polarization of the scalar (linear polarized) Bessel beam was experimentally measured as it propagated through the obstruction. The results are shown in the first row of Fig. 3(a). As can be seen, as the scalar Bessel beam propagated through the obstruction from (I) to (IV), its intensity self-heals as is expected \cite{Review1,Review2, Review3, Dudley1}. Next, the self-healing of the intensities and spatially inhomogeneous states of polarizations of the radialyl and azimuthally polarized vector Bessel beams were experimentally measured as they propagated through the obstruction. The results are shown in the second and third rows of Fig. 3(a), respectively. As the radially and azimuthally polarized vector Bessel beam propagated through the obstruction from (I) to (IV), their intensities self-healed similar to the scalar Bessel beam. As can be seen, similar to their intensities, the spatially inhomogeneous states of polarization of the radially and azimuthally polarized vector Bessel beams also self-healed. This is the salient result of this Letter. There is qualitative agreement between the spatially inhomogeneous states of polarization of the vector Bessel beams when they are unobstructed at (I) and when they self-heal at (IV). 

Also, the self-healing of the intensities and spatially inhomogeneous states of polarization of the radially and azimuthally polarized vector Bessel beams were experimentally measured as they propagated through a disparate obstruction. The results are shown in Fig. 3(b). As shown in Fig. 3(b), the size and position of this obstruction was chosen such that it obstructed a larger portion of the vector Bessel beams near their center. The obstruction is outlined by a dashed white line. As can be seen, the intensities and the spatially inhomogeneous states of polarization of the radially and azimuthally polarized vector Bessel beams again self-heal in the presence of the larger obstruction. Similar to the first obstruction, there is qualitative agreement between the spatially inhomogeneous states of polarization of the vector Bessel beams when they are unobstructed at (I) and when they self-heal at (IV).  

Similar, to scalar Bessel beams, the self-healing of vector Bessel beams can be understood via geometric optics, i.e., the interference of conical rays in the shadow region of the obstruction, as shown in Fig. 1(e). It is a well-known that the distance in which a Bessel beam is able to reform is given by \cite{Litvin}:
\begin{eqnarray}
z_{min} \approx \frac{a k}{2k_z}
\end{eqnarray}
where $a$ is the width of the obstruction and $k$ and $k_z$ are the wave-vector and longitudinal wave-vector, respectively. Eq. 6 illustrates that the distance in which a Bessel beam self-heals is dependent on the size and position of the obstruction as well as the opening angle of the cone on which the wave-vectors of the Bessel beam propagate. Here it is assumed that the input field is larger than the obstruction.

There is a relationship between light's space and polarization degrees of freedom when light is scattered by an obstruction, e.g. Rayleigh or Mie particles\cite{Milione5}. There are comparable relationships  in multimode optical fiber \cite{Milione6, Milione7}. In this respect, a more detailed theoretical analysis of self-healing of the spatially inhomogeneous states of polarization of radially and azimuthally polarized vector Bessel beams, analogous to that of scalar Bessel beams \cite{Litvin}, is the subject of future work. Nonetheless, as vector Bessel beams possess the properties of Bessel beams and a vector beams, they may have applications in, for example, optical trapping, where self-healing and ``vectorness" are both needed, e.g. it may be possible to the improve the axial and transverse stiffness of a tractor beam when using a vector Bessel beam.

In conclusion, we experimentally measured the self-healing of the spatially inhomogeneous states of polarization of radially and azimuthally polarized vector Bessel beams. Radially and azimuthally polarized vector Bessel beams were generated via a digital version of DurninÕs method, using an SLM in concert with a liquid crystal $q$-plate. As a proof of principle, their intensities and spatially inhomogeneous states of polarization were measured using Stokes polarimetry as they propagated through two disparate obstructions. It was found, similar to their intensities, the spatially inhomogeneous states of polarization of a radially and azimuthally polarized vector Bessel beams self-heal. Similar, to scalar Bessel beams, the self-healing of vector Bessel beams can be understood via geometric optics, i.e., the interference of conical rays in the shadow region of the obstruction. The self-healing of vector Bessel beams may have applications in, for example, optical trapping. 

While there are extensive studies on self-healing of scalar Bessel beams\cite{Review1,Review2, Review3, Dudley1}, there are limited studies on self-healing  of vector Bessel beams \cite{Hasman, Vyas, He,Wu}, particularly with respect to their spatially inhomogeneous states of polarization. Previous work only measured the propagation and self-healing of the \textit{intensities} of vector Bessel beams. To our knowledge, this is the first experimental measurement of the self-healing of the spatially inhomogeneous states of \textit{polarization} of radially and azimuthally polarized vector Bessel beams. Future work includes experimentally measuring the self-healing of the spatially inhomogeneous states of polarization of other types of vector Bessel beams such as Full Poincare beams \cite{Amber}. In contrast to radially and azimuthally polarized light beams, Full Poincare beams experience non-trivial dynamics as they propagate \cite{Amber2, Milione3}.

Acknowledgments: We thank S. Slussarenko and L. Marrucci for the $q$-plates. GM acknowledges helpful discussions with D. Grier and D. Ruffner, and support from AFOSR Grant. No.  No. 47221-00-01, ARO Grant. No. 52759-PH-H, NSF GRFP Grant. No. 40017-00-04, and Corning, Inc. E.K. acknowledges the support of the Canada Excellence Research Chairs (CERC) Program.

\end{document}